\documentclass[10pt, amsmath, amssymb, amsfonts]{article}

\usepackage{bm,graphicx}
\usepackage[english]{babel}
\usepackage{epsfig, amssymb}

\newcommand{\finpr}{\hfill $\square$ \vspace{2mm}}
\def\be{\begin{eqnarray}}
\def\ee{\end{eqnarray}}
\def\bee{\begin{eqnarray*}}
\def\eee{\end{eqnarray*}}
\newtheorem{thm}{Theorem}
\newtheorem{cor}{Corollary}
\newtheorem{lem}{Lemma}

\newtheorem{defn}{Definition}

\begin{document}

\title{\bf  Simulating quantum computers with \\ probabilistic methods}

\author{Maarten Van den  Nest \\ \\
{\normalsize\it Max-Planck-Institut f\"ur Quantenoptik,} \\ {\normalsize\it  Hans-Kopfermann-Stra\ss e 1, D-85748 Garching, Germany. }}

\maketitle

\begin{abstract}
We investigate the boundary between classical and quantum computational power. This work consists of two parts. First we develop new classical simulation algorithms that are centered on sampling methods. Using these techniques we generate new classes of classically simulatable quantum circuits where standard techniques relying on the exact computation of measurement probabilities fail to provide efficient simulations. For example,  we show how various concatenations of matchgate, Toffoli, Clifford, bounded-depth, Fourier transform and other  circuits are classically simulatable. We also prove that sparse quantum circuits as well as circuits composed of CNOT and $\exp[{i\theta X}]$ gates can be simulated classically. In a second part, we apply our results to the simulation of quantum algorithms. It is shown that a recent quantum algorithm, concerned with the estimation of Potts model partition functions, can be simulated efficiently classically. Finally, we show that the exponential speed-ups of Simon's and Shor's algorithms crucially depend on the very last stage in these algorithms, dealing with the classical postprocessing of the measurement outcomes. Specifically, we prove that both algorithms would be classically simulatable if the function classically computed in this step had a sufficiently peaked Fourier spectrum.
\end{abstract}

\section{Introduction}

What is the power of quantum computers compared to classical ones?  Understanding this fundamental but difficult question is one of the great challenges in the field of quantum computation.

A fruitful approach to tackle this  problem is to study classes of quantum computations that do \emph{not} offer any computational benefits over classical computation. Indeed, such investigations shed light on the essential features of quantum mechanics that are responsible for quantum computational power. At the same time, understanding which classes of quantum computations can be simulated classically provides useful insights in the difficult task of constructing  novel quantum algorithms, potentially yielding indications on where to look for new algorithmic primitives.

In recent years several non-trivial classes of quantum computations have been identified for which an efficient classical simulation  can be achieved. For example, certain  computations are classically simulatable due to the absence of high amounts of entanglement (quantified appropriately in terms of suitable entanglement measures) \cite{Jo02,Vi03,Yo06,Jo06,Va06}. Other well known results are the Gottesman-Knill theorem \cite{Go98, De03, Aa04, Cl07, Va08} and the classical simulation of matchgate circuits \cite{Va02, Di04, Br09, Jo08, Jo09}. The latter two classes of results provide key illustrations of the fascinating and puzzling relation between classical and quantum computational power, as they e.g. regard computations that may exhibit large degrees of entanglement, interference, superposition, etc.---i.e. the ingredients that supposedly provide QC with its increased power---but which nevertheless cannot achieve any computational speed-up over classical computers.

A common element in many existing classical simulation results and methods is the notion of classical simulation that is, sometimes implicitly, adopted in these works. When a quantum computation is to be simulated classically, the goal may be to either classically compute measurement probabilities (or expectation values) with high precision in poly-time (``strong simulation''), \emph{or} to classically sample in poly-time from the resulting output probability distribution (``weak simulation''). Given the intrinsic probabilistic nature of quantum mechanics, it is readily motivated that weak simulation is  the more natural notion of what a classical simulation should constitute. Furthermore, one may easily construct examples of quantum circuit classes for which strong simulation is intractable whereas weak simulation is achieved by elementary sampling methods (see e.g. \cite{Va08})---hence showing that a gap between strong and weak simulations manifests itself already in elementary scenarios. The latter gap moreover highlights that any serious attempt to compare classical with quantum  computational power should not be based on strong simulation methods.

In spite of these basic and well-known insights, the majority of existing results on classical simulation of QC regard the strong variant, and weak simulation techniques seem to date largely unexplored. The goal of the present work is to develop new classical simulation algorithms that are based on sampling methods and  to therewith initiate an investigation of the potential of weak simulation of quantum computation. Next we state more precisely the contributions of this work.

\section{Statement of results}

\subsection*{Classical simulation of QC with probabilistic methods}

In a first part of the paper, we develop tools to investigate weak classical simulation of QC. A central ingredient in our analysis will be a certain class of quantum states, called here \emph{computationally tractable states} (CT states). Colloquially speaking, a state is CT if it is possible to classically simulate computational basis measurements on $|\psi\rangle$ \emph{and} if the coefficients of $|\psi\rangle$ in this basis can be efficiently computed. As we will see, many important state families---matrix product states, stabilizer states, states generated by poly-size matchgate circuits, and several others---turn out to be CT. A second element will be the notion of efficiently computable sparse operators (ECS). An $n$-qubit operation is ECS if its matrix representation in the standard basis has at most poly$(n)$ nonzero entries per row and per column, and if these entries can be determined efficiently.  For example, all Pauli products, $k$-local operators with $k=O(\log n)$, as well as operators that can be written as poly-size circuits of Toffoli gates, are ECS. We will prove the following result.

\begin{thm}\label{thm_CT}
Consider a poly-size quantum circuit acting on a state $|\psi\rangle$ and followed by measurement of an observable $O$. If $|\psi\rangle$ is computationally tractable and if $U^{\dagger} O U$ is efficiently computable sparse, then this quantum computation can be simulated classically.
\end{thm}
An immediate remark to be made is that the unitary operation $U$ itself is \emph{not} required to be sparse---only its action on  $O$ is to yield an ECS operation, which is a significantly distinct requirement. For example, if $U$ is a poly-size circuit consisting of nearest neighbor matchgates---which is generally not sparse at all---then $U^{\dagger}ZU$ is a linear combination of poly$(n)$ Pauli products, which is an ECS operation.

Theorem \ref{thm_CT}  identifies a general scenario in which quantum circuits can be simulated efficiently classically. This result turns out to be rather versatile and will be useful in a number of contexts. In this work we highlight the following particular applications  (however, it is likely that this result has applications beyond the ones considered here):

\begin{figure}\hspace{0.5cm}
{\includegraphics[width=6.5cm]{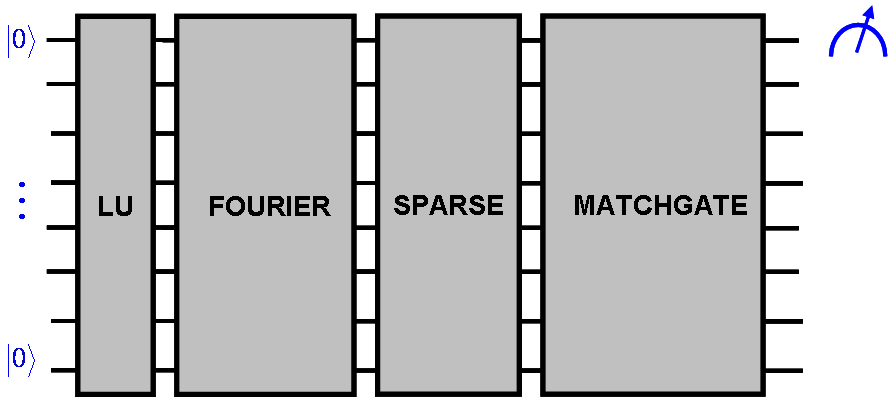}}\hspace{0.5cm}{\includegraphics[width=6.5cm]{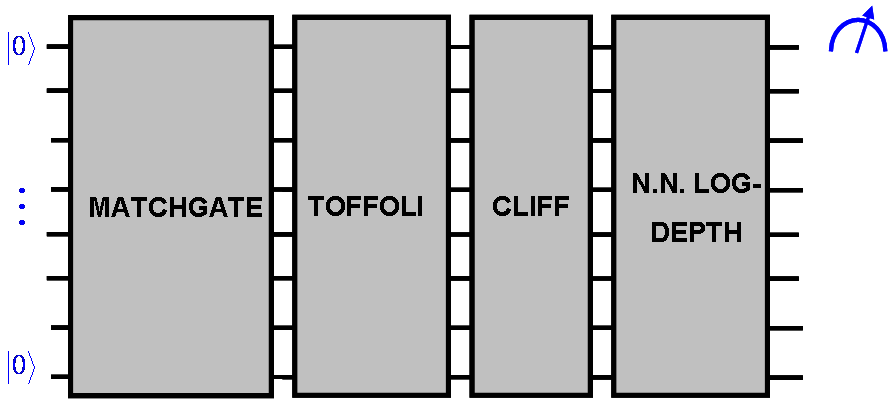}}
\caption[]{\label{Fig:examples} The above concatenated quantum circuits can be efficiently simulated classically via an application of theorem \ref{thm_CT}. See section \ref{sect_comp} for a discussion of these examples.}
\end{figure}

\begin{itemize}
\item {\bf Sparse circuits.} A simple instance of theorem \ref{thm_CT} is obtained by considering a product input state (which is trivially CT) and the $Z$ observable on, say, the first qubit, and by letting the circuit $U$ itself be an ECS operation (in which case $U^{\dagger} ZU$ is ECS as well).  Then, by virtue of theorem \ref{thm_CT}, the resulting quantum computation can be simulated classically. In fact, one can immediately extend this result by composing $m$ efficiently computable $s$-sparse\footnote{An operator is $s$-sparse if its standard basis matrix representation has at most $s$ nonzero entries per row and per column.} unitary operations with $s^m=$ poly$(n)$. Then the overall circuit will still be ECS, as can easily be verified, and thus can be simulated classically due to theorem \ref{thm_CT}.

    Sparse unitary operations are of interest because they highlight the role of \emph{interference} in quantum computation, as opposed to \emph{entanglement}. In particular, sparse operations may produce highly entangled states but the interference exhibited in any sparse unitary evolution is always limited. As we will show, this absence of high degrees  of interference can be exploited to construct an efficient classical simulation algorithm, in spite of the potentially complex entangled states produced throughout the computation. This provides (yet another) illustration that the presence of entanglement is by no means sufficient to guarantee quantum computational speed-ups.  Sparse operations furthermore provide examples of a class of QCs where weak classical simulation is efficiently possible, whereas strong simulation is intractable ($\#$P-hard). In other words, adopting the notion of weak simulation constitutes a necessary ingredient in the simulation of sparse circuits whereas strong simulation methods such as e.g. tensor contraction schemes cannot (unless $\#$P $=$ P) yield an efficient classical simulation.

\item {\bf Composability.} Instead of letting $|\psi\rangle$ be a simple product input state, we may also consider more complicated CT states which are e.g. the result of an earlier quantum computation, i.e. $|\psi\rangle = U'|\psi_{\mbox{\scriptsize{in}}}\rangle$ for some simple (e.g. standard basis) input $|\psi_{\mbox{\scriptsize{in}}}\rangle$. As long as $|\psi\rangle$ is CT \emph{and} subsequently a circuit $U$ is applied followed by measurement of $O$ such that $U^{\dagger}OU$ is ECS, the overall quantum circuit $UU'$, acting on $|\psi_{\mbox{\scriptsize{in}}}\rangle$ and followed by measurement of $O$, can be simulated classically by theorem \ref{thm_CT}. One hence arrives at a criterion to asses when the \emph{concatenation} of two quantum circuits can be simulated classically.

\begin{figure}
\hspace{5cm}{\includegraphics[width=6cm]{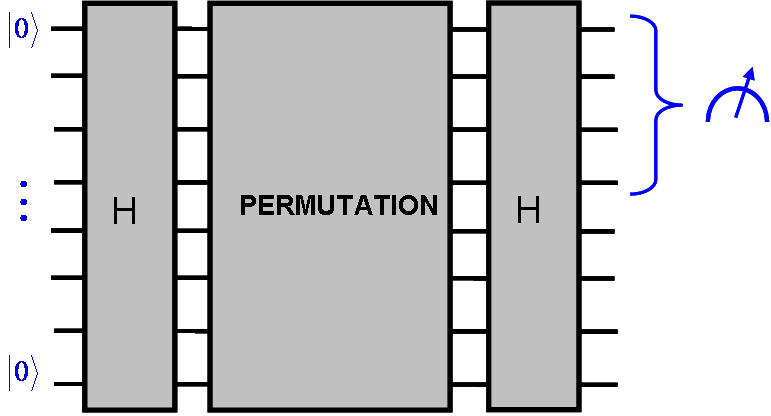}}
\caption[]{\label{Fig:structure_shor} Both the  factoring algorithm and Simon's algorithm can be implemented by a circuit with the following structure. The first and third round in the circuit consist of collections of Hadamard operations applied to certain subsets of the qubits; the second round is a unitary operation that acts as a permutation on the computational basis. The circuit is followed by a $\{|0\rangle, |1\rangle\}$ measurement of a subset of the qubits. The algorithm concludes with classical postprocessing of the measurement results.}
\end{figure}

    Since the majority of existing efficiently simulatable circuits  turn out to generate CT states when acting on suitable inputs \emph{and} as at the same time many simulatable operations yield ECS operations when acting on suitable observables, the above composability result is applicable to a wide variety of settings. In particular, this result applies to Clifford operations, matchgate circuits, bounded-depth circuits, classical circuits, bounded-treewidth circuits, the quantum Fourier transform, and others. This leads to sometimes surprising examples of concatenated circuits that can be simulated classically (cf. Fig. \ref{Fig:examples}).
    As illustrated in these examples, the concatenation of simulatable blocks of very different nature may remain efficiently simulatable classically (consider e.g. the concatenation of a Clifford with a matchgate circuit).

    It is interesting to compare the examples in Fig. \ref{Fig:examples} to powerful quantum algorithms such as Simon's and Shor's. Strikingly, the latter algorithms are implemented with particularly simple circuitry---arguably even simpler than the classically simulatable circuits displayed in Fig. \ref{Fig:examples}. In particular,  it is known that both the factoring algorithm  and Simon's algorithm can be efficiently implemented by a circuit with the very simple structure of Fig. \ref{Fig:structure_shor} \cite{Ki95,Si97}.  Intriguingly,  this circuit is the composition of only three blocks, each of which is elementary. Nevertheless, our simulation techniques \emph{cannot} be successfully be applied to yield an efficient classical simulation of this circuit class. In the second part of this work we investigate the hardness of simulating these circuits and, by extension, Simon's and Shor's algorithms,  in more detail.

\item {\bf CNOT-$e^{i\theta X}$ circuits.} As a further application of theorem \ref{thm_CT}, we will show that poly-size circuits composed of CNOT and $e^{i\theta X}$ gates, acting on  product inputs and followed by measurement of $Z$ on any single qubit, can be simulated classically.
    This result is of interest since it is known that  CNOT together with any single \emph{real} one-qubit gate $V$ such that $V^2$ is not basis-preserving, is \emph{universal} for quantum computation \cite{Sh02}. In contrast to this, here it is found that there is a class of non-trivial \emph{complex} gates $e^{i\theta X}$ that can be added to the CNOT gate while retaining efficient classical simulation.

    The above result is also interesting from a conceptual point of view. In particular, its proof will follow from a variant of theorem \ref{thm_CT} where states $|\psi\rangle$ and operations $U^{\dagger}OU$ are considered that are CT, resp. ECS, with respect to bases other than the standard basis. Letting $|\psi\rangle$ be a product input and $U$ a poly-size circuit composed of CNOT and $e^{i\theta X}$ gates, it will be shown that $|\psi\rangle$ and $U^{\dagger}Z_1U$ are CT, resp. ECS, \emph{with respect tho the $\{|\pm\rangle\}$ basis of $X$ eigenstates}. Hence, viewing the entire computation in this basis and applying theorem \ref{thm_CT} shows that classical simulation is efficiently possible. In contrast, a direct application of theorem \ref{thm_CT}, i.e. with respect to the standard basis, is not possible as $U^{\dagger}Z_1U$ is generally not ECS w.r.t this basis.

\end{itemize}

\subsection*{Classical simulation of quantum algorithms}\label{sect_intro_algo}

In a second part of the paper, the above results  are applied  in the context of quantum algorithms. Depending on the case at hand, the goal will be to either show that certain algorithms can be simulated classically or to deepen our insight into why certain algorithms achieve exponential (oracle) speed-ups over classical computation. We will analyze three different quantum algorithms:
\begin{itemize}
\item a  quantum algorithm to estimate partition functions of classical lattice models \cite{Ar08};

\item a general class of quantum algorithms containing the Deutsch-Jozsa algorithm \cite{De92};

\item Simon's algorithm \cite{Si97}.
\end{itemize}
The first two classes of quantum algorithms will be proved to be  classically simulatable  using the methods developed in this paper.
We refer to the relevant sections in the text for a discussion. For the time being, we limit ourselves to discussing our results in the context of Simon's algorithm, which we consider the most interesting application.

Recall that in  Simon's problem one has oracle access to a function $f:\{0, 1\}^n\to\{0, 1\}^n$; it is promised that there exists an unknown $n$-bit string $a$ such that $f(x)= f(y)$ if and only if $y=x+a$ (addition modulo 2). The goal is to find $a$. Classically one needs at least $O(2^{\frac{n}{2}})$ queries,  whereas a quantum computer can solve the problem with $O(n)$ queries---i.e. Simon's algorithm achieves an exponential oracle separation between  BQP and BPP. In spite of its computational power, Simon's algorithm is implemented with very simple circuitry, as displayed in Fig. \ref{Fig:structure_shor}. What are the essential ingredients responsible for the power of this algorithm?

In standard considerations, the interplay between the Fourier transform (i.e. the second layer of Hadamards in Fig. \ref{Fig:structure_shor}) and the oracle $f$ is emphasized. After the oracle is applied, the system is in the state $\sum |x\rangle|f(x)\rangle$. The Fourier transform then creates interference in the system and ``picks out'' the relevant computational basis states, such that a subsequent measurement of the system yields the desired information about the unknown bit-string $a$. This rather delicate relation between oracle and Fourier transform is usually considered to be among the main origins of the hardness of classically simulating Simon's algorithm. In this work we will show that this point of view is not the end of the story: in particular, we will find that the interplay between the same Fourier transform and the function computed during the round of  \emph{classical postprocessing} is an equally important element in the speed-up achieved by the algorithm. Specifically, we will prove the following result.

\begin{thm}[rough version]\label{thm_postproc} Consider a quantum circuit displaying the structure as depicted in Fig. \ref{Fig:structure_shor}. If the function computed in the round of classical postprocessing is promised to have a sufficiently ``peaked'' Fourier spectrum, then the entire circuit can be simulated efficiently classically, independent of the specific forms of the other rounds.
\end{thm}
Thus, if the final classical round in Simon's algorithm happened to  regard a function with sufficiently peaked fourier spectrum, then the entire quantum computation could be simulated efficiently ---independent of the details of e.g. the oracle $f$ computed in an earlier stage of the computation, and independent of e.g. the entanglement produced by the quantum circuit. This result hence exposes the double role played by the Fourier transform, which is to act appropriately on \emph{both} the oracle $f$ \emph{and} the function computed in the postprocessing, in order to achieve a quantum speed-up. These observations highlight that the power of a quantum algorithm can only be understood by taking the entire computation into account including the classical postprocessing round, even though the latter may at first sight look rather innocuous. Indeed, note that---strikingly---in Simon's algorithm this round `only' involves solving a simple system of \emph{linear} equations over $\mathbb{Z}_2$! Nevertheless, this simple classical computation is associated with a function having a very \emph{flat} spectrum (as we will see), hence ensuring the exponential speed up achieved by Simon's algorithm.

{\bf Remark:} in the formulation of theorem 2, no knowledge of the Fourier spectrum of the function in question is assumed, except the promise that this spectrum is ``peaked''. Using remarkable results of Boolean learning theory, enough information of the spectrum can be efficiently reconstructed in order to achieve the poly-time classical simulation as stated in the theorem.\hfill $\diamond$

Finally, also the factoring algorithm can be implemented with a circuit displaying the  structure of Fig. \ref{Fig:structure_shor}. Therefore, the classical postprocessing plays a similar crucial role also in this algorithm. As the technical considerations in Simon's algorithm are more transparent than in Shor's, here we will focus on the former---keeping in mind that our conclusions also apply to the latter.

\subsection*{Matchgate circuits and poly-time classical computation}
Somewhat unrelated to the above context, we prove a ``byproduct result'' that we find noteworthy. We will arrive at a complexity-theoretic result regarding the computational power of matchgate circuits. Roughly speaking, we will show the following (see theorem \ref{thm_mg} for a precise statement):
\begin{itemize}
\item[] \emph{The class of functions that can be efficiently computed by nearest-neighbor matchgate circuits is strictly contained within P}.
\end{itemize}
Perhaps the most interesting aspect regarding this result here is its proof method. Surprisingly, the result will be obtained by combining the classical query lower bound of Simon's problem with our theorem \ref{thm_CT}.  In particular, we will show that if the class of matchgate-computable  functions comprised all of P, then a quantum algorithm for Simon's problem would exist which  turns out to be efficiently simulatable classically (using theorem \ref{thm_CT}). Hence an efficient \emph{classical} algorithm would exist which solves Simon's problem with poly$(n)$ classical oracle queries, yielding a contradiction. Remark that it is striking how utterly unrelated matchgate circuits and Simon's problem seem at first sight!

\subsubsection*{Some conventions}

In this paper, when we refer to a quantum circuit, we will always implicitly mean a uniformly generated family of quantum circuits. Further, by observable we mean any Hermitian operator $O$ with $\| O\|\leq 1$, where $\|\cdot\|$ denotes the spectral norm. When a measurement of an observable is considered at the end of a quantum circuit, we will always implicitly assume that this regards an observable that can be measured efficiently. The notion of `simulation' will be synonymous to `classical simulation'. The notion `efficient' will be synonymous  to `in polynomial time'.
For clarity, all results are stated in terms of qubit systems, but generalizations to arbitrary finite-dimensional quantum systems are immediate. Our standard notation for the computational basis of an $n$-qubit system will be $\{|x\rangle\}$, where $x=(x_1, \dots, x_n)$ ranges over all $n$-bit strings and $|x\rangle = |x_1\rangle\otimes \dots\otimes|x_n\rangle$.

\section{Classical simulation of quantum computation}\label{sect_simulation}

In this section we discuss the definition of classical simulation that will be adopted in the present work.  Suppose that  an $n$-qubit poly-size quantum circuit produces an output state $|\psi_{\mbox{\scriptsize{out}}}\rangle$ and is followed by a measurement of an observable $O$, assuming that $O$ can efficiently be measured. Then, repeating the computation  $K=$ poly$(n)$ times, recording the measurement outcome $o_i$ in each run (i.e. each $o_i$ is one of the eigenvalues of $O$) one obtains an estimate $\sigma= K^{-1} \sum_{i=1}^K o_i$ of the expectation value $\langle O\rangle = \langle\psi_{\mbox{\scriptsize{out}}} |O|\psi_{\mbox{\scriptsize{out}}}\rangle$. The accuracy of this approximation is dictated by the Chernoff-Hoeffding bound (we refer to the Appendix for a statement and  discussion of this bound). In particular, this bound implies the following: for every $\epsilon = p(n)^{-1}$, where $p(n)$ represents an arbitrary polynomial in $n$, there exists a $K$ that scales as a suitable polynomial in $n$ such that the inequality $| \sigma - \langle O\rangle |\leq \epsilon$ holds with a probability that is exponentially (in $n$) close to 1. In other words, by taking poly$(n)$ runs of the computation---and this is all that is allowed in an efficient quantum computation---it is possible to estimate $\langle O\rangle$ with an error that scales as an arbitrary inverse polynomial in $n$. We denote this type of estimate as an approximation with `polynomial accuracy' or a `polynomial approximation'. Note that a polynomial approximation achieves an estimate of $\langle O\rangle$ up to $O(\log n)$ significant bits.

The above method hence represents an efficient quantum algorithm to  estimate $\langle O\rangle$ with polynomial  accuracy with a success probability that lies exponentially close to 1. We now say that this quantum algorithm can be efficiently simulated classically if there exists an efficient classical algorithm to provide a polynomial approximation of $\langle O\rangle$, again with a probability that lies exponentially close to 1. That is, we require the classical simulation algorithm to approximate $\langle O\rangle$ in poly-time \emph{with the same accuracy that is achieved by the quantum algorithm}. This notion of simulation is sometimes called \emph{weak simulation}. The latter is to be regarded as opposed to the much more stringent requirement of \emph{strong} simulation, where it is asked to construct a classical algorithm to approximate $\langle O\rangle$ in poly$(m, n)$ time up to $m$ significant bits (i.e. with exponential precision).

Note that the notion of weak simulation is more true to the concept of  what a classical simulation actually  constitutes since, colloquially speaking, it requires the classical simulation to achieve `the same result' as the quantum algorithm.  In contrast, in the strong scenario one is asked to construct an efficient classical algorithm that approximates $\langle O\rangle$ far more accurate than the quantum algorithm itself could generally achieve in polynomial time. Even though it has been realized previously that the weak variant is a valid and natural notion of classical simulation of QC (see e.g. \cite{Jo02, Jo08}), it seems that this notion is to date largely unexplored. In particular, the vast majority of  classical simulation results  use the strong variant. In \cite{Va08} it was pointed out  that there exists simple examples of quantum circuits for which weak classical simulation is possible with elementary methods, whereas strong simulation of the same circuits is a $\#$P-hard problem and hence intractable. This highlights the presence of a significant gap between strong and weak simulation.

{\bf Remark:} When the notion of polynomial approximation is used in the following, we will always mean a polynomial approximation which is achieved with a probability that is exponentially close to one. \hfill $\diamond$

\section{Computationally tractable states}\label{sect_CT}
The objective of this section is to develop the notion of computationally tractable (CT) states and to prove theorem \ref{thm_CT}. To do this, first we first define CT states and discuss some of their elementary properties; this is done in section \ref{sect_CT_def}. In section \ref{sect_basis_preserving} we consider basis-preserving operations, which are identified as a class of operations that map CT states to CT states. In section \ref{sect_sparse} we consider sparse operations; the main technical contribution in this section is theorem \ref{thm_main} regarding the efficient classical estimation of matrix elements $\langle\varphi|A|\psi\rangle$, where $|\psi\rangle$ and $|\varphi\rangle$ are computationally tractable and $A$ is an (efficiently computable) sparse operation. This theorem will immediately lead to the proof of theorem \ref{thm_CT}.

\subsection{Definition of CT states}\label{sect_CT_def}

Throughout this paper, we will deal with  $n$-qubit state families $\{|\psi_n\rangle: n=1, 2, \dots\}$, where $|\psi_n\rangle$ is an $n$-qubit state. When considering such a state family $\{|\psi_n\rangle\}$, we will mostly refer to a single state $|\psi_n\rangle\equiv|\psi\rangle$ with the silent assumption that this actually denotes a family. We  now consider the following definition.
\begin{defn}
An $n$-qubit state  $|\psi\rangle$ is called `computationally tractable' (CT) if the following conditions hold:
\begin{itemize}
\item[(a)]   it is possible to sample in poly$(n)$ time with classical means from the probability distribution Prob$(x)=|\langle x|\psi\rangle|^2$ on the set of $n$-bit strings $x$, and \item[(b)] upon input of any bit string $x$, the coefficient $\langle x|\psi\rangle$ can be computed in poly$(n)$ time on a classical computer.
\end{itemize}
\end{defn}
For convenience, in (b) we require the coefficients $\langle x|\psi\rangle$ to be computable with perfect precision, a notion which may lead to rather pathological situations when e.g. irrational numbers are involved. The results in this paper can however straightforwardly be generalized to the case where $\langle x|\psi\rangle$ can be computed efficiently with exponential precision, i.e. up to $m$ significant bits in poly$(n, m)$ time. As in the present work the distinction between these two types of accuracies is not essential (in contrast to the distinction between polynomial and exponential precision, which \emph{is} crucial), for clarity we state all results  w.r.t. the notion of perfect accuracy. Also in other places in the text where we refer to `perfect accuracy', the results in question  immediately generalize to the case of exponential precision.

Note that (a) and (b) are highly dependent on the classical description of the state $|\psi\rangle$  that is provided. Therefore, strictly speaking it would be more precise to call a state $|\psi\rangle$ CT \emph{relative to this classical description}. In this paper we will only encounter situations where each state has a natural  (efficient) description  that will be obvious from the context. It will always be assumed that  this particular description is provided.  For example, the classical description of a state generated by a poly-size quantum circuit acting on, say, the all-zeroes input, will always be assumed to be the circuit that generates the state. As another example, for every complete product state $|\psi\rangle=|\psi_1\rangle\otimes\dots\otimes|\psi_n\rangle$ we will assume $|\psi\rangle$ to be specified in terms of the `obvious' description of $|\psi\rangle$ consisting if the  $2n$ complex coefficients $\langle 0|\psi_i\rangle$ and $\langle 1|\psi_i\rangle$.

Even though conditions (a) and (b) are similar in nature, we provide evidence that these conditions are incomparable. In particular, the following complexity theoretic argument implies that it is highly likely that there exists states satisfying (b) but not (a). Consider any efficiently computable function $f:\{0, 1\}^n\to \{0, 1\}$ for which it is promised that there exists a unique $x_0$ such that $f(x_0)=1$, and define the $n$-qubit state $|\psi\rangle =\sum_x f(x)|x\rangle = |x_0\rangle$. Note that the state $|\psi\rangle$ satisfies condition (b). Assuming that (b) implies (a), it follows that it is possible to efficiently sample from the distribution $\{|\langle x|\psi\rangle|^2\}$. But this distribution assigns a zero probability to each bit string $x$ except $x_0$, which has unit probability. Hence, the possibility of efficiently sampling from this distribution implies that $x_0$ can be determined efficiently. Regarding $f$ as a verifier circuit for an NP problem, it would immediately follow that every problem in NP with a unique witness is in P. This last property is not likely to be true \cite{Va85}.

Next we state a useful sufficient  (but not necessary) criterion to assess whether condition (a) holds for a given state. To state this result, we need the following notation. For an $n$-qubit state $|\psi\rangle$, let  $p_{S, y}(|\psi\rangle)\equiv p_{S, y}$ denote the probability of obtaining the bit string $y=(y_i : i\in S)$ as an outcome when measuring the qubits in the set $S\subseteq\{1, \dots, n\}$. We can then state the following lemma; a proof  can be found in e.g. \cite{Va02}.
\begin{lem}\label{lem_marginal}
Let $|\psi\rangle$ be an $n$-qubit state.  Suppose that, on input of an arbitrary $S$ and $y$, the probability $p_{S, y}$ can be computed in poly$(n)$ time. Then it is possible to sample in poly$(n)$ time from the probability distribution $\{|\langle x|\psi\rangle|^2\}$.
\end{lem}

Several important state families turn out to be computationally tractable, as illustrated next.

\begin{itemize}
\item {\bf Examples of computationally tractable states:} \item[--] Product states are trivially CT. \item[--] Every state of  the form $|\psi\rangle\propto \sum_x e^{i \theta(x)} |x\rangle$, where the sum is over all $n$-bit strings $x$ and where $x\to\theta(x)\in \mathbb{R}$ represents an arbitrary efficiently computable function, is trivially CT. Every state obtained by applying a poly-size circuit family consisting of { Toffoli} gates to an arbitrary product state is computationally tractable as well, as can easily be proved (this property will also follow from lemma \ref{basis-preserving}). \item[--] Every matrix product state (MPS) of polynomial bond dimension is CT.  A state $|\psi\rangle$ is an MPS  of poly bond dimension if there exist $2n$ $N\times N$  matrices $A_i[0], A_i[1]$ with $N=$ poly$(n)$ such that $\langle x|\psi\rangle$ $=$ $\mbox{ Tr}(A_1[x_1]\dots A_n[x_n])$, for every $n$-bit string $x=(x_1, \dots, x_n)$. Property (b) follows immediately from this definition.  Property (a) holds since the conditions of lemma \ref{lem_marginal} are satisfied for all MPS of polynomial bond dimension \cite{Pe07}. Tree tensor states \cite{Sh06} are generalizations of MPS with similar properties and are also computationally tractable. \item[--] A Clifford circuit is a quantum circuit composed of Hadamard, CNOT and PHASE gates, where  PHASE $=$ diag$(1, i)$. An $n$-qubit stabilizer state is any state that is generated by applying a poly-size Clifford circuit to the state $|0\rangle^n$. Every stabilizer state is a CT state. Property (a) is  the content of the Gottesman-Knill theorem \cite{Go98}. Property (b) is proved in \cite{De03} (see also \cite{Va08}).

\item[--] A (unitary, two-qubit) matchgate $G$ is any two-qubit gate of the form \be G = \left[\begin{array}{cccc} a & & & b \\ & u&v&\\&x&y& \\ c & & & d\end{array} \right],\quad A =\left[\begin{array}{cc} a & b\\ c & d\end{array} \right], \quad B=\left[\begin{array}{cc} u & v\\ x & y\end{array} \right], \ee where $A, B\in SU(2)$. Every state obtained by applying a poly-size matchgate circuit to a computational basis state, where all gates are restricted to act on nearest neighbors (assuming a one-dimensional ordering of the qubits) is a computationally tractable state. Properties (a) and (b) are proved in \cite{Va02}.

\item[--] Any $n$-qubit state that is obtained by applying the quantum Fourier transform (over the integers modulo $2^n$) to an arbitrary product state, is a CT state. See e.g. \cite{Br07} for a simple proof of this property (see also  \cite{Yo07, Ah06} for related results).

\item[--] We briefly mention a general class of classical simulation results related to efficient tensor contraction schemes. This approach relies on the topology of (a graph associated with) the quantum circuit in question.  If this topology displays a sufficiently tree-like structure (quantified in terms of the graph invariant \emph{tree-width}) then classical simulation of such circuits can be achieved \cite{Ma05}.  It can be shown that the output states of quantum circuits with logarithmically scaling tree-width (acting on product input states), are CT states; the proof essentially contained in \cite{Ma05} and is omitted here (see also \cite{Jo06} for related work).
\end{itemize}

\subsection{Basis-preserving  operations}\label{sect_basis_preserving}

Next we investigate which operations map the family of CT states to itself. In this context,  the operations that preserve the computational basis play an important role. An $n$-qubit operation $M$ is called `basis-preserving' if  every computational basis state $|x\rangle$ is mapped to $M|x\rangle=\gamma_x|{\pi(x)}\rangle$, for some permutation $\pi$ of the set of $n$-bit strings and some complex  $\gamma_x$. The operation $M$ is efficiently computable if  the functions $x\to \gamma_x$, $x\to\pi(x)$ and $x\to \pi^{-1}(x)$ can be evaluated in poly$(n)$ time. For example, every Pauli product \cite{foot3} is efficiently computable basis-preserving, as well as every operation of the form $O = \sum_x (-1)^{f(x)} |x\rangle\langle x|$, where $f:\{0, 1\}^n\to \{0, 1\}$ is an efficiently computable function. Also every poly-size circuit composed of elementary basis-preserving gates (e.g. Toffoli gates, diagonal gates) is efficiently computable basis-preserving.

The relevance of efficiently computable basis-preserving \emph{unitary} operations in the present context is that these operations preserve the class of CT states:

\begin{lem} \label{basis-preserving} If $|\psi\rangle$  is a computationally tractable $n$-qubit state and if $M$ is an efficiently computable unitary basis-preserving operation, then $|\psi'\rangle = M|\psi\rangle$ is again computationally tractable.
\end{lem}
{\it Proof: } Let the permutation $\pi$ and the coefficients $\gamma_x$ be defined as above. Note that $|\gamma_x|=1$ for every $x$ since $M$ is unitary. The coefficients of $|\psi'\rangle$ are given by $\langle x|\psi'\rangle = \gamma_{\pi^{-1}(x)} \langle {\pi^{-1}(x)}|\psi\rangle$. Property (b) now follows immediately from the properties that $M$ is efficiently computable and that $|\psi\rangle$ is CT. To show (a), we have to find an efficient classical method to sample from the probability distribution defined by Prob$(x)=|\langle x|\psi'\rangle|^2 = |\langle {\pi^{-1}(x)}|\psi\rangle|^2$. To do so, consider the following procedure. First sample from the distribution $\{|\langle  y|\psi\rangle|^2\}$, yielding a bit string $y$ with probability $|\langle  y|\psi\rangle|^2$, and subsequently output the bit string $x:=\pi(y)$. This procedure is efficient since $|\psi\rangle$ is CT and $y\to\pi(y)$ is efficiently computable. Moreover, every bit string $x$ is generated with probability $|\langle {\pi^{-1}(x)}|\psi\rangle|^2$ as desired. \finpr

Note that the basis-preserving operation $M$ may drastically change the entanglement properties of $|\psi\rangle$. Consider e.g. the case where $|\psi\rangle$ is a complete product state and $M$ a poly-size circuit  of CPHASE and/or Toffoli operations, yielding a state $|\psi'\rangle$ that may be highly entangled. Nevertheless, both $|\psi\rangle$ and $|\psi'\rangle$ are CT and equal up to a basis-preserving operation.

\subsection{Sparse operations}\label{sect_sparse}

Next we consider sparse operations.  Such operations are sufficiently close to basis-preserving operations that their action on CT states remains manageable. An $n$-qubit operation $A$ is  $s$-sparse if for every basis state $|x\rangle$, each of the vectors $A|x\rangle$ and $A^T|x\rangle$ is a linear combination of at most $s$ computational basis states. The quantity $s$ is called the sparseness of $A$. We will consider $n$-qubit operations $A$ (both unitary operations and observables)  with sparseness $s\leq $ poly$(n)$, which will simply be called `sparse operations'. Note that the notion of sparseness is defined w.r.t. to the number of nonzero entries per row/column and \emph{not} the total number of nonzero entries in the matrix, the latter not being required to be small. In particular, a sparse $n$-qubit operation generically has a total number of nonzero entries that scales \emph{exponentially} with $n$.

For every $s$-sparse $n$-qubit operation $A$,  define $2s$ functions $\alpha_i: \{0, 1\}^n\to \mathbb{C}$ and $r_i: \{0, 1\}^n\to\{0, 1\}^n$ ($i=1, \dots, s$) as follows: the $n$-bit string $r_i(x)$ is defined to be the  row index  of $A$ associated with the $i$-th non-zero entry in the column indexed by $x$ (when traversing this column from top to bottom), if an $i$-th nonzero entry exists within this column; we denote  this entry by $\alpha_i(x)$. If an $i$-th nonzero entry does not exist in this column, then $r_i(x)$ is set to be the all-zeroes string and $\alpha_i(x)$ is set to zero. With the above definitions, one simply has \be\label{sparse} A|x\rangle = \alpha_1(x) | {r_1(x)}\rangle +\dots + \alpha_s(x) | {r_s(x)}\rangle.\ee Similar definitions can be given regarding the rows of $A$, leading to $2s$ functions $\beta_i: \{0, 1\}^n\to \mathbb{C}$ and $c_i: \{0, 1\}^n\to\{0, 1\}^n$ ($i=1, \dots, s$) that are the natural counterparts of the $\alpha_i$ and $r_i$, respectively.

A sparse $n$-qubit operation $A$ is efficiently column-computable if, on input of an arbitrary $n$-bit string $x$, it is possible to list the (at most $s=$ poly$(n)$) nonzero entries within the column of $A$ indexed by $x$ together with the row indices associated with each of these non-zero entries, all in poly$(n)$ time. Equivalently, $A$ is efficiently column-computable if it is possible to compute the $2s$ quantities $\alpha_i(x)$ and $r_i(x)$ ($i=1, \dots, s$) in poly-time.  The operation $A$ is called efficiently row-computable if $A^T$ is efficiently column-computable. Finally, $A$ is called efficiently computable if it is both efficiently row- and column-computable.  All efficiently computable sparse unitary operations can be implemented efficiently on a quantum computer \cite{Jor09}. In this paper we will only consider sparse operations that are efficiently computable.

The following are some  examples of efficiently computable sparse operations.

\begin{itemize}
\item {\bf  Examples of efficiently computable sparse (ECS) operations:}

\item[--] Every efficiently computable basis-preserving operation is ECS.

\item[--] Every $d$-qubit gate $G$ acting within an $n$-qubit circuit, represented by the matrix $G\otimes I$ where $I$ denotes the identity acting on $n-d$ qubits, is $2^d$-sparse. If $d=O(\log n)$ then such an operation is ECS.

\item[--] Every operation that is a linear combination of poly$(n)$ ECS operations, is ECS. It follows that every operator $H=\sum_{i=1}^m H_i$ which is a sum of $m=$ poly$(n)$ $d$-local observables $O_i$ (with $d=O(\log n)$) is ECS. This means that observables such as Hamiltonians and correlation operators are typically ECS.

\item[--] Let $U$ represent an $n$-qubit poly-size circuit  of basis-preserving elementary gates (e.g. Toffoli, CNOT, PHASE, CPHASE, etc.), interspersed with $k$ gates $V_1, \dots, V_k$ at arbitrary places in the circuit, each of which acts on at most $d$ qubits. It is required that $kd = O(\log n)$; otherwise the $V_i$ are arbitrary. Then $U$ is  ECS. To see this, expand each gate $V_i$ as a linear combination of $4^d$ Pauli products and note that every Pauli product is efficiently computable basis-preserving.  Consequently, $U$ can be written as a linear combination of $4^{dk}=$ poly$(n)$ efficiently computable basis-preserving operations, showing that $U$ is ECS.

\item[--] ECS operations often arise in the context of quantum algorithms, related e.g. to unitary group representations; see e.g. \cite{Jor09} and references within.

\end{itemize}

We are now in a position to state the following  result, which constitutes the main technical ingredient in this work regarding the use of sampling techniques in classical simulation.

\begin{thm}\label{thm_main}
Let $|\psi\rangle$ and $|\varphi\rangle$ be CT $n$-qubit states and let $A$ be an efficiently computable sparse (not necessarily unitary) $n$-qubit operation with $\| A\|\leq 1$. Then there exists an efficient classical algorithm to approximate $\langle\varphi|A|\psi\rangle$ with polynomial accuracy.
\end{thm}
Note that theorem \ref{thm_CT} immediately follows from theorem \ref{thm_main}.  Before proving this result in its most general form, as a warm-up we prove a special instance, taking $A$ to be the identity. Hence, we are concerned with the estimation of overlaps between CT states.  This special case is proved beforehand  to illustrate the sampling methods used in this work, without the more technically involved arguments required in the proof of theorem \ref{thm_main}. Thus, we set out to prove the following property, formulated in terms of a lemma.

\begin{lem}\label{thm_overlaps} Let $|\psi\rangle$ and $|\varphi\rangle$ be two CT $n$-qubit states. Then there exists an efficient classical algorithm to approximate $\langle\varphi|\psi\rangle$ with polynomial accuracy.
\end{lem}
{\it Proof: }   Denote $p_x:= |\langle x|\psi\rangle|^2$ and $q_x:= |\langle x|\varphi\rangle|^2$. Since $|\psi\rangle$ and $|\varphi\rangle$ are CT states, it is possible to sample efficiently from the probability distributions $\{p_x\}$ and $\{q_x\}$. Define the function $\delta:\{0, 1\}^n\to \{0, 1\}$ by $\delta(x)= 1$ if $p_x\geq q_{x}$ and $\delta(x) =0$ otherwise, for every $n$-bit string $x$, and define $\epsilon = 1-\delta$. Then $\delta$ and $\epsilon$ can be evaluated efficiently since $p_x$ and $q_x$ can be efficiently evaluated by assumption (b) in the definition of CT states. The overlap $\langle\varphi|\psi\rangle$ is therefore equal to \be \langle\varphi|\psi\rangle = \sum \langle\varphi|x\rangle\langle x|\psi\rangle \delta(x) + \sum \langle\varphi|x\rangle\langle x|\psi\rangle \epsilon(x),\ee where the sums are over all $n$-bit strings $x$. Defining the functions $F$ and $G$ by \be F(x)= \frac{\langle\varphi|x\rangle\langle x|\psi\rangle}{p_x} \ \delta(x), \quad G(x)=\frac{\langle\varphi|x\rangle\langle x|\psi\rangle}{q_x} \ \epsilon(x), \ee we have $\langle\varphi|\psi\rangle = \langle F\rangle + \langle G\rangle$ where $\langle F\rangle= \sum p_x F(x)$ and $\langle G\rangle = \sum q_{x} G(x)$. It follows from assumption (b) in
the definition of CT states that $F$ and $G$ can be efficiently evaluated. Furthermore, both $|F(x)|$ and $|G(x)|$ are not greater than 1. It thus follows from the Chernoff-Hoeffding bound that both $\langle F\rangle$ and $\langle G\rangle$ can be approximated efficiently with polynomial accuracy. This implies that $\langle\varphi|\psi\rangle$ can be estimated with polynomial accuracy as well. This completes the proof.
\finpr

Lemma \ref{thm_overlaps} shows that the overlap $\langle\varphi|\psi\rangle$, representing a `joint' property of the states $|\psi\rangle$ and $|\varphi\rangle$, may be estimated efficiently classically even when only an efficient simulation of quantum processes resulting in $|\psi\rangle$ and $|\varphi\rangle$ \emph{individually} is available---in particular, the techniques leading to the proofs  of (a)-(b) (cf. definition of CT states) for $|\psi\rangle$ and $|\varphi\rangle$, may be completely different. For example, the overlap between a matrix product state and a stabilizer state can be estimated efficiently classically with polynomial accuracy, even though such states are CT due to very different argumentations.

\

We are now in a position to prove theorem \ref{thm_main}.

{\bf Proof of theorem \ref{thm_main}:} It is sufficient to prove the result for CT states $|\psi\rangle$ and $|\varphi\rangle$. Let $s=$ poly$(n)$ denote the sparseness of $A$. Using the notation of (\ref{sparse}), we have $\langle\varphi|A|\psi\rangle = \sum_{i=1}^n \sigma_i$, where we denote \be \sigma_i:= \sum_x \alpha_i(x)\langle\varphi| {r_i(x)}\rangle\langle x|\psi\rangle.\ee Note that $|\alpha_i(x)|\leq 1$. It is sufficient to prove that each of the $s$ quantities $\sigma_i$ can be estimated efficiently with polynomial accuracy, for then also $\sum_{i=1}^s \sigma_i$ can be estimated with polynomial accuracy as $s=$ poly$(n)$. To do so, write $p_x:= |\langle x|\psi\rangle|^2$ and $q_x:= |\langle x|\varphi\rangle|^2$. Define a function $\delta_i$ by $\delta_i(x)= 1$ if $p_x\geq q_{r_i(x)}$ and $\delta_i(x) =0$ otherwise, for every $n$-bit string $x$, and define $\epsilon_i = 1-\delta_i$. Then $\delta_i$ and $\epsilon_i$ can be evaluated efficiently since $|\psi\rangle$ and $|\varphi\rangle$ are CT and $A$ is ECS. We split $\sigma_i$ in two parts by inserting $\delta_i(x) + \epsilon_i(x)=1$: \be\label{sigma} \sigma_i = \sum \langle\varphi| {r_i(x)}\rangle\langle x|\psi\rangle \alpha_i(x)\delta_i(x) +\sum \langle\varphi| {r_i(x)}\rangle\langle x|\psi\rangle \alpha_i(x)\epsilon_i(x).\ee The function $F_i$ defined by \be\label{F_i} F_i(x)= \frac{\langle\varphi| {r_i(x)}\rangle\langle x|\psi\rangle}{p_x} \ \alpha_i(x)\delta_i(x)\ee is efficiently computable and satisfies $|F_i(x)|\leq 1$ for every $x$. The first term in the r.h.s. of (\ref{sigma}) is hence equal to $\langle F_i\rangle = \sum p_x F_i(x)$, which can be estimated to polynomial accuracy efficiently due to the Chernoff-Hoeffding bound. To estimate the second term in the r.h.s. of (\ref{sigma}), one needs to be careful since the function $r_i$ may not be injective. We proceed as follows. Define the following function $G_i$: \be\label{sum_G} G_i(y)= \sum_{x:\ r_i(x)=y \mbox{ \scriptsize{ and }} \alpha_i(x)\neq 0} \frac{\langle\varphi| y\rangle\langle x|\psi\rangle}{q_{y}}\ \alpha_i(x)\epsilon_i(x)\ee with the additional convention that $G_i(y)$ is zero if there are no $x$ such that $r_i(x)=y$ and $\alpha_i(x)\neq 0$. With this definition, the second term in the r.h.s. of (\ref{sigma}) is  equal to $\langle G_i\rangle = \sum_y q_y G_i(y)$. We now make the following claims. {\it Claim 1}:  the function $G_i$ is efficiently computable; and {\it Claim 2}: $|G_i(y)|\leq s$ for every $y$. A proof of claims 1 and 2 implies that $\langle G_i\rangle$ can be estimated in poly-time with polynomial accuracy due to the Chernoff-Hoeffding bound. But then also $\sigma_i$ can be estimated efficiently, thus completing the proof.

We now prove Claim 1. Since $A$ is $s$-sparse, every row $y$ has at most $s$ non-zero entries. Equivalently, the following set contains at most $s$ strings $x$: \be\label{setofstrings} \{ x: \exists j\in\{1, \dots, s\} \mbox{ s.t. } y=r_j(x) \mbox{ and } \alpha_j(x)\neq 0\}.\ee Hence, a fortiori, for every fixed $i$ there are at most $s$ different $x$ such that $r_i(x)=y$ and $\alpha_i(x)\neq 0$. Moreover, given an arbitrary $y$ it is possible to efficiently determine all these $x$'s and the corresponding coefficients $\alpha_i(x)$. This is done in two steps: first, since $A$ is efficiently (row-)computable, given a row index $y$ it is possible to compute all (at most $s$) strings $x$ in the set (\ref{setofstrings}) in poly-time; second, for all those $x$ one computes $r_i(x)$ and $\alpha_i(x)$---this is possible in poly-time since $A$ is efficiently column-computable---and verifies whether $r_i(x)$ is equal to $y$; those $x$ for which $r_i(x)=y$ are kept, the others discarded.

It follows that $G_i(y)$ is a sum of at most $s=$ poly$(n)$ terms, each of which is efficiently computable. Thus, Claim 1 is proved. Moreover, Claim 2 now immediately follows as well, since the modulus of every term in the sum (\ref{sum_G}) is smaller than one and there are at most $s$ terms in the sum. This proves theorem \ref{thm_main}.\finpr

{\bf Remark: poly-ECS operations.---} In the definition of ECS operations and in the subsequent statement of theorem \ref{thm_main}, we have required that the non-zero entries of $A$ can be computed efficiently with perfect precision. Theorem \ref{thm_main} also holds for sparse operations where, instead, these coefficients can be estimated efficiently with \emph{polynomial} accuracy, which is a significant relaxation.  Call an $n$-qubit operation $A$ $(\|A\|\leq 1)$ \emph{poly-ECS} if it is sparse, and if (i) on input of an arbitrary column index $x$, it is possible to determine in poly-time all those row indices $y$ such that $\langle y|A|x\rangle \neq 0$ and if the corresponding nonzero entries  $\langle y|A|x\rangle$ can be estimated in poly-time with polynomial accuracy, and (ii) similarly for the row indices $y$.
Theorem \ref{thm_main} then also holds for poly-ECS operations. The proof is completely analogous to the above proof of theorem \ref{thm_main}. The only difference is that now the functions $F_i(x)$ and $G_i(x)$ can no longer be computed exactly, but only with polynomial accuracy. However, this suffices to invoke the Chernoff-Hoeffding bound (cf. the Appendix). This remark will play an important role in the discussion of Simon's algorithm i.e. in the proof of theorem \ref{thm_postproc}.\hfill $\diamond$

\

We conclude this section with  two corollaries of theorem \ref{thm_main}. Corollary \ref{cor_montecarlo} shows that expectation values of local observables can be estimated efficiently classically for every CT state. This result may potentially be of use in e.g. variational Monte Carlo studies of strongly correlated systems (this is work in progress). Corollary \ref{cor_partial} will be of use when we discuss the Deutsch-Jozsa algorithm in section \ref{sect_DJ}.

\begin{cor}\label{cor_montecarlo}
Let $|\psi\rangle$ be an $n$-qubit CT state and let $O$ be a $d$-local observable with $d=O(\log n)$ and $\|O\|\leq 1$. Then there exists an efficient classical algorithm to estimate $\langle \psi|O|\psi\rangle$ with polynomial accuracy.
\end{cor}
{\it Proof:} this result follows immediately from theorem \ref{thm_main} since every $d$-local $O$ with $d=O(\log n)$ is ECS. Here we provide a short alternative proof that does not require the formalism used in the proof of theorem \ref{thm_main}. Every observable $O$ of the form considered can be written as a linear combination of $N=$ poly$(n)$ Pauli operators: $O= \sum_{i=1}^N a_i P_i$, with $|a_i|\leq 1$. Consequently, \be \langle O\rangle := \langle \psi|O|\psi\rangle = \sum a_i \langle\psi|P_i|\psi\rangle.\ee As each $P_i$ is an efficiently computable basis-preserving unitary operation, each state $P_i|\psi\rangle$ is CT due to lemma \ref{basis-preserving}. Invoking lemma \ref{thm_overlaps}, the overlap between $P_i|\psi\rangle$ and $|\psi\rangle$ can be estimated classically with polynomial accuracy. Hence, also $\langle O\rangle$ can be estimated classically with polynomial accuracy. This proves the result. \finpr

\begin{cor}\label{cor_partial}
Let $|\psi\rangle$ and $|\varphi\rangle$ be CT $n$-qubit states, let $|\xi\rangle$ and $|\chi\rangle$ be CT $k$-qubit states (with $k\leq n$) and let $A$ and $B$ be efficiently computable sparse $n$-qubit operations with $\| A\|, \| B\|\leq 1$. Then there exists an efficient classical algorithm to approximate $\langle\varphi|A[|\xi\rangle\langle\chi|\otimes I]B|\psi\rangle$ with polynomial accuracy.
\end{cor}
{\it Proof:} The proof uses a technique related to the SWAP test. Denote $|\psi'\rangle:= B|\psi\rangle$ and $|\varphi'\rangle:= A^{\dagger}|\varphi\rangle$ (which are potentially unnormalized states) and consider the following identity: \be\label{swap} \langle\varphi'|[|\xi\rangle\langle\chi|\otimes I]|\psi'\rangle = [\langle\chi|\langle\varphi'|] U_{\mbox{\scriptsize{SWAP}}} [|\xi\rangle|\psi'\rangle],\ee where the unitary operator $U_{\mbox{\scriptsize{SWAP}}}$ swaps qubit $i$ with qubit $i+k$, for every $i=1, \dots, k$. The identity (\ref{swap}) can easily be verified. Hence, we have \be \langle\varphi|A[|\xi\rangle\langle\chi|\otimes I]B|\psi\rangle  = [\langle\chi|\langle\varphi|] [I\otimes A] U_{\mbox{\scriptsize{SWAP}}} [I\otimes B][|\xi\rangle|\psi\rangle].\ee Note that the $(k+n)$-qubit states $|\xi\rangle|\psi\rangle$ and $|\chi\rangle|\varphi\rangle$ are CT. Moreover, it can easily be verified that $U_{\mbox{\scriptsize{SWAP}}}$ is ECS. This implies that the operation $[I\otimes A] U_{\mbox{\scriptsize{SWAP}}} [I\otimes B]$ is ECS as well, being a product of three ECS operations. Theorem \ref{thm_main} can now be applied. \finpr

Note that, as a special case of this last result, it follows that partial overlaps $\langle\varphi|[|\xi\rangle\langle\chi|\otimes I]|\psi\rangle$ between CT states can be estimated efficiently classically.

\section{Applications of theorem \ref{thm_CT}}

Next we discuss three applications of theorem \ref{thm_CT} as announced in the introduction. These applications regard sparse circuits, composability, and CNOT-$e^{i\theta X}$ circuits.

\subsection{Classical simulation of sparse circuits}

The following is a formal statement of the classical simulation of sparse circuits which was announced in the introduction.

\begin{cor}\label{thm_sparse} Let $U$ be a circuit composed of $m$ efficiently computable $s$-sparse unitary operations with $s^m=$ poly$(n)$. The circuit acts on an arbitrary product input state and is followed by a $Z$ measurement of, say, the first qubit. Then this quantum computation can be simulated efficiently classically.\end{cor}
{\it Proof:} Let $|\psi\rangle$ denote the product input state and let $Z_1$ denote the $Z$ observable acting on the first qubit. The expectation value of $Z_1$ is given by $\langle Z_1\rangle = \langle\psi| U^{\dagger}ZU|\psi\rangle$. Note that $U$ is ECS due to the restrictions on $s$ and $m$; but then also the observable $O:=U^{\dagger}ZU$ is ECS, being a product of three ECS operations. Moreover, $|\psi\rangle$ is a product state and hence CT. Theorem \ref{thm_CT} can now be applied.\finpr

As briefly alluded to in the introduction, sparse operations highlight the role of  interference---as opposed to entanglement---in quantum computation. Note that sparse operations may generically produce highly entangled states. Consider e.g. the simple case where the input is $|+\rangle^n$ and the entire circuit $U$ is composed of poly$(n)$ CPHASE gates (which are basis-preserving gates and thus particularly simple examples of sparse operations). With such circuits, it is possible to efficiently generate e.g. the highly entangled cluster states \cite{He06}.
On the other hand, if a sparse operation $U$ acts on a state $|\psi\rangle$ then each coefficient of $U|\psi\rangle$ in the standard basis is a linear combination of at most poly$(n)$ coefficients of $|\psi\rangle$. Hence, the ``interference'' in the process $|\psi\rangle\to U |\psi\rangle$ is limited (we use the notion of interference in a colloquial sense and do not adopt any technical definition). Corollary \ref{thm_sparse} states that quantum computational processes where the interference is ``small'' in this sense, cannot offer any speed-up compared to classical computers, in spite of the high degrees of entanglement that may be generated throughout the computation. Corollary \ref{thm_sparse} may thus be regarded as complementary to a class of results stating  that quantum computations that generate low amounts of entanglement (quantified appropriately) can be classically simulated efficiently (see e.g. \cite{Jo02,Vi03, Yo06, Jo06, Va06}).

Finally, note that in corollary \ref{thm_sparse} one cannot hope for an improvement of the bound $s^m=$ poly$(n)$ to e.g. $m=$ poly$(n)$ and $s$ constant (unless BQP = BPP) since \emph{every} poly-size quantum circuit is a product of $m=$ poly$(n)$ single- and two-qubit gates, each of which is an $s$-sparse operation with $s$ constant.

\subsection{Composability}\label{sect_comp}

Theorem \ref{thm_CT} immediately leads to a criterion to assess when the composition of two quantum circuits can be simulated classically. Formally, we have:

\begin{cor}\label{thm_comp1}
Consider poly-size $n$-qubit quantum circuits $U_1$ and $U_2$, an input state $|\psi_{\mbox{\scriptsize{in}}}\rangle$ and an observable $O$ such that: (i) the state $U_1|\psi_{\mbox{\scriptsize{in}}}\rangle$ is computationally tractable and (ii) the operation $U_2^{\dagger}OU_2$ is efficiently computable sparse. Then the circuit $U=U_2U_1$, acting on $|\psi_{\mbox{\scriptsize{in}}}\rangle$ and followed by measurement of $O$, can be simulated efficiently classically.
\end{cor}
Next we provide some illustrations of this result. First we provide some examples of pairs $(U, O)$ such that $U^{\dagger}OU$ is ECS. All circuit families $U$ below are poly-size.

\begin{itemize}
\item {\bf Examples of pairs $(U,O)$ where $U^{\dagger}OU$ is ECS:}

\item[--] Let $U$ be a circuit of constant depth and let the observable $O$ act nontrivially on $O(\log n)$ qubits. Then $U^{\dagger}OU$ also acts nontrivially on  $O(\log n)$ qubits and is hence an ECS observable.

\item[--] Let $U$ represent a Clifford  circuit and let $O$ be any observable that is a linear combination of $N=$ poly$(n)$ Pauli products: $O=\sum_{i=1}^N a_i P^i$ with $|a_i|\leq 1$ and $P^i$ Pauli operators. Then $U^{\dagger}OU$ is again a linear combination of $N$ Pauli products, and hence ECS.

\item[--]Let  $U$ be a circuit composed of nearest-neighbor matchgates and let $Z_1$ denote the Pauli $Z$ operation acting on the first qubit. Then $U$ maps $Z_1$ (under conjugation) to a linear combination of poly$(n)$ Pauli products (see e.g. \cite{Jo08}), which is an ECS operation.
\end{itemize}

Next we explicitly describe two concatenated circuits that can be simulated efficiently using our results; see also Fig \ref{Fig:examples}. In both examples, the circuit acts on the all-zeroes computational basis state and is followed by measurement of $Z$ on the first qubit.

\begin{itemize}

\item {\bf Examples of corollary \ref{thm_comp1}:}

\item[--] Consider a quantum circuit $V=V_4V_3V_2V_1$ where $V_1$ is an arbitrary local unitary operation, $V_2$ represents the quantum Fourier transform (over $\mathbb{Z}_{2^n}$), $V_3$ is an arbitrary efficiently computable sparse unitary, and $V_4$ is an arbitrary poly-size (nearest-neighbor) matchgate circuit. Then this circuit can
be simulated efficiently classically due to corollary \ref{thm_comp1}. In particular, we show that corollary \ref{thm_comp1} can be applied by taking $U_1\equiv V_2V_1$ and $U_2\equiv V_4V_3$. To see this, note first that $V_2V_1$ acting on the input yields a CT state. Further, $(V_4V_3)^{\dagger}Z(V_4V_3)$ is ECS: indeed, $V_4^{\dagger}ZV_4$ is a sum of poly$(n)$ Pauli products and hence ECS, and thus $(V_4V_3)^{\dagger}Z(V_4V_3)$ is ECS as well, being a product of three ECS operations. Corollary \ref{thm_comp1} can now be applied.

\item[--] Consider a quantum circuit $V=V_4V_3V_2V_1$ where $V_1$ is an arbitrary poly-size matchgate circuit, $V_2$ is a poly-size circuit of Toffoli gates,   $V_3$ is an arbitrary poly-size Clifford circuit and $V_4$ is an arbitrary log-depth circuit consisting of nearest-neighbor gates. We
show that corollary \ref{thm_comp1} can be applied by taking $U_1\equiv V_1$ and $U_2\equiv V_4V_3V_2$. To see this, note first that $V_1$ acting on the input yields a CT state. Further, $(V_4V_3V_2)^{\dagger}Z(V_4V_3V_2)$ is ECS: $V_4^{\dagger}ZV_4$ acts nontrivially on $O(\log n)$ qubits and is hence is a linear combination of poly$(n)$ Pauli products; but then also $V_3^{\dagger}V_4^{\dagger}ZV_4V_3$ is a linear combination of poly$(n)$ Pauli products (and hence ECS) since $V_3$ is a Clifford operation; finally, it follows that $(V_4V_3V_2)^{\dagger}Z(V_4V_3V_2)$ is ECS as this operation is a product of three ECS operations. Corollary \ref{thm_comp1} thus again yields the desired result.
\end{itemize}

Several other examples of the above nature can easily be generated.

\subsection{Rotated bases and CNOT-$e^{i\theta X}$ circuits}\label{sect_comp_B_general}

In our definition of computationally tractable states and sparse operations, as well as in the resulting theorem \ref{thm_CT}, we have singled out a particular basis---i.e. the computational basis. Note, however, that in the vast majority of all arguments we have never relied on the specific form of this basis. Therefore, we may consider a generalized definition of CT states, sparse operations, etc., stated \emph{relative to a arbitrary basis ${\cal B}$}, and carry out an analogous program as done so far, leading a much broader class of  results. Results such as theorem \ref{thm_CT} can  be transferred in an obvious  way, and will be omitted. Here we limit ourselves to discussing an example that can be understood using this generalized notion of CT states. This example regards the simulation of circuits composed of CNOTs and $e^{i\theta X}$ gates. Other examples of similar nature can easily be constructed.

Let ${\cal B}= \{|b_x\rangle\}$ denote the $|\pm\rangle$ product basis,  defined by $|b_x\rangle \propto \bigotimes_{i=1}^n [|0\rangle + (-1)^{x_i}|1\rangle]$ for every $n$-bit string $x=(x_1, \dots, x_n)$. A state is called `computationally tractable in the basis ${\cal B}$' if  it is possible to sample in poly$(n)$ time with classical means from the probability distribution Prob$(x)=|\langle b_x|\psi\rangle|^2$, and if the coefficients $\langle b_x|\psi\rangle$ can be computed in poly$(n)$ time classically. It is clear that  $|\psi\rangle$ is CT in ${\cal B}$ iff $H^{\otimes n}|\psi\rangle$ is CT in the computational basis. For example, it can easily be shown that every stabilizer state, as well as any MPS $|\psi\rangle$ is computationally tractable in the $|\pm\rangle$-basis ${\cal B}$ as $H^{\otimes n}|\psi\rangle$  is in both cases CT in the computational basis.

Similarly, the notion of ECS operations w.r.t. ${\cal B}$ is defined in the natural way.  Obviously, $A$ is ECS w.r.t. ${\cal B}$ iff $H^{\otimes n}A H^{\otimes n}$ is ECS in the computational basis.  For example, let $U$ denote an arbitrary poly-size $n$-qubit circuit composed of CNOT and $e^{i\theta X}$ gates, where $\theta$ may be any (real) angle. Whereas $U$ is generally \emph{not} ECS in the computational basis, this circuit  is \emph{always} ECS in the $|\pm\rangle$  basis ${\cal B}$. This can be seen as follows.   Let CNOT$_{ab}$ denote a CNOT gate with control $a$ and target $b$. One then has the pair of identities \be H^{\otimes 2} \mbox{CNOT}_{ab} H^{\otimes 2} = \mbox{ CNOT}_{ba} \quad \mbox{and} \quad He^{i\theta X} H = e^{i\theta Z},\ee both of which are easily verified. These identities imply that $M:=H^{\otimes n} U H^{\otimes n}$ is a poly-size circuit consisting entirely of CNOT and $ e^{i\theta Z}$ gates and is thus ECS (even basis-preserving) in the computational basis. This shows that $U$ is ECS in the $|\pm\rangle$ product basis.

One can now consider a generalized form of theorem \ref{thm_CT}, now stated relative to the $|\pm\rangle$ basis (or any other basis):

\begin{itemize}
\item[] {\bf Theorem 1'} {\it
Let $|\psi_{\mbox{\scriptsize{in}}}\rangle$ be an $n$-qubit state, let $U$ denote a poly-size $n$-qubit circuit and let $O$ denote an observable. If $|\psi\rangle$ is CT in ${\cal B}$ and if $U^{\dagger} O U$ is ECS in ${\cal B}$, then the circuit $U$, acting on $|\psi_{\mbox{\scriptsize{in}}}\rangle$ and followed by measurement of $O$, can be simulated efficiently classically.}
\end{itemize}
Now consider a CNOT-$e^{i\theta X}$ circuit $U$ as above. The circuit $U$ acts on an arbitrary product input $|\alpha\rangle$ and is followed by measurement of $Z_1$. We now claim that this computation can be simulated efficiently classically, using the above variant of theorem \ref{thm_CT}. To see this, first note that $|\alpha\rangle$ is CT in ${\cal B}$. Second, $O:=U^{\dagger}Z_1U$ is ECS in ${\cal B}$: to show this, note that $H^{\otimes n}O H^{\otimes n} = M^{\dagger} X_1 M.$ Here, as before, $M:=H^{\otimes n} U_2 H^{\otimes n}$ is a poly-size circuit consisting entirely of CNOT and $ e^{i\theta Z}$ gates, and $X_1$ denotes the Pauli $X$ operation acting on the first qubit.  The operation $M^{\dagger} X_1 M$ is basis-preserving in the computational basis, hence $O = H^{\otimes n}[M^{\dagger} X_1 M]H^{\otimes n}$ is basis-preserving in ${\cal B}$. This proves the claim; note that we have hence proved:

\begin{cor}\label{thm_CNOT_X} Every poly-size circuit composed of CNOT and $e^{i\theta X}$ gates (for arbitrary real $\theta$), acting on an arbitrary product input and followed by measurement of $Z_1$, can be simulated efficiently classically.
    \end{cor}

\section{Simulating quantum algorithms}\label{sect_alg}

In this section we apply our results in the context of quantum algorithms. The idea is to consider e.g. theorems \ref{thm_CT} and \ref{thm_main} and corollary \ref{cor_partial} as a collection of `tests' that every quantum algorithm claiming to achieve an exponential speed-up needs to pass. We will consider the three classes of algorithms  mentioned in the introduction.

\subsection{Potts models}\label{sect_Potts}

Here we point out that a recently proposed quantum algorithm \cite{Ar08}, concerned with estimating partition functions of classical spin systems such as the Potts model,   can be simulated efficiently classically. Letting ${\cal Z}$ denote the Potts model partition function defined on some (arbitrary) lattice, the quantum algorithm in \cite{Ar08} provides a polynomial approximation of the quantity ${\cal Z}/\Delta$. Here $\Delta$ denotes a particular, easy-to-compute normalization factor that depends on the couplings of the model (see \cite{Ar08}, Cor. 5.9, for the precise form of $\Delta$); $\Delta$ is sometimes called the `approximation scale' of the algorithm. On the other hand, in \cite{Va07} mappings were established which allow to express the same quantity ${\cal Z}/\Delta$ as the overlap between a suitable product state $|\alpha\rangle$ and  stabilizer state $|\psi\rangle$: ${\cal Z}/{\Delta} = \langle\alpha|\psi\rangle$. Note that both stabilizer states and product states are CT (see section \ref{sect_CT}). Using theorem \ref{thm_main} (in fact: the special instance $A=I$ of lemma \ref{thm_overlaps}, dealing with overlaps between CT states),  we find that overlaps between stabilizer states and product states can also efficiently be estimated with polynomial accuracy with \emph{classical} methods. Hence, the quantity ${\cal Z}/{\Delta}$ can also be estimated with polynomial accuracy in poly-time using classical means, showing that the quantum algorithm in question can be simulated efficiently classically.

We emphasize that the work \cite{Ar08} contains several quantum algorithms besides the partition function algorithm focused on here (in particular, the latter does not constitute the main result of \cite{Ar08}), including algorithms for BQP-complete problems, to which our classical simulation techniques do not apply.

\subsection{Deutsch-Jozsa}\label{sect_DJ}
An application of corollary  \ref{cor_partial} is found by considering the Deutsch-Jozsa (DJ) algorithm \cite{De92}. Recall that in the DJ problem one considers a black-box function $f:\{0, 1\}^n\to \{0, 1\}$ which is promised to be either constant or balanced \cite{foot4''}. The task is to determine which possibility holds. Classically, any deterministic solution to the problem requires exponentially many oracle calls, whereas a randomized classical algorithm can solve the DJ problem with exponentially small probability of failure using $O(n)$ queries.  The DJ quantum algorithm constitutes a deterministic solution to the problem using a single query of the oracle.

Thus, it is well known that DJ can be simulated classically when an exponentially small probability of failure is allowed. Here we will reproduce this result, showing that it immediately follows from corollary  \ref{cor_partial}. Moreover, we will find that a large class of generalizations (to be specified below) can  be efficiently simulated as well. The argument is very general and mainly regards the \emph{structure} of the involved circuits.

Going through the steps in the DJ algorithm, it is easily verified that DJ  is implemented by a circuit belonging to the following general class  (the system is initialized in the state $|0\rangle^n$):

\begin{itemize}
\item[] {\bf Round  1}: apply a local unitary operation $V_1$;

{\bf Round  2}: apply an ECS operation $V_2$;

{\bf Round  3}: apply another local unitary operation $V_3$ ;

{\bf Round  4:}  measure the observable $O=|0\rangle\langle 0|^k\otimes I$, for some $k\leq n$.
\end{itemize}
Using corollary  \ref{cor_partial}, we now immediately find that such a computation can be simulated efficiently classically. Indeed, the state obtained after Round 1 is a  a product state and hence CT. Moreover, the operation in round 2 is efficiently computable sparse. Finally, the observable $O':= V_3^{\dagger}OV_3$ has the form $|\gamma\rangle\langle\gamma|\otimes I$ for some $k$-qubit product---and hence CT---state $|\gamma\rangle$. Corollary \ref{cor_partial} can now immediately be applied.

Note that, in the argument, the specific form of the function $f$ (computed in Round 2) is completely irrelevant. This shows that the lack in computational power of the DJ algorithm is a \emph{structural} feature of the circuit. In particular, this computational weakness cannot be overcome by e.g. changing the form of the oracle, but must involve a more drastic alteration of the circuit structure.

\subsection{Simon's algorithm} \label{sect_simon}

Lastly, we consider Simon's algorithm \cite{Si97}.  As this algorithm has the admirable feature of being a very simple quantum algorithm that nevertheless achieves an exponential speed-up, it is an ideal candidate to compare quantum and classical computational power. Simon's algorithm is worth investigating from a number of angles. As a comprehensive study would lead us too far, here we single out one particular aspect, namely the surpising role of the round of classical postprocessing in the algorithm taking place \emph{after} the measurement.  We will show that this seemingly innocuous round of classical computation plays a rather determining role in the performance of the algorithm.

We first give a short review of Simon's algorithm in section \ref{sect_review_simon}. In section \ref{sect_fourier} we take small detour, discussing aspects of Fourier analysis of Boolean functions, which will be necessary to prove theorem \ref{thm_postproc}; the latter is done in section \ref{sect_proof_thm2}.

\subsubsection{Review of Simon's algorithm}\label{sect_review_simon}

Here we will focus on a decision problem version of Simon's problem, where it is asked to determine the $i$-th bit $a_i$ of the unknown string $a$ for some $i$. We will fix $i=1$ in the following for concreteness.

Simon's quantum algorithm consists of the following steps. There are two registers, each  consisting of $n$ qubits, each initially prepared in the state $|0\rangle^n$. First a Hadamard operation is applied to every qubit in the first register. Second, the  oracle operator $U_f$ is applied, yielding $\sum_x |x\rangle|f(x)\rangle$. Third, again a Hadamard operation is applied to every qubit in the first register. This yields a state of the form $|\psi_{\mbox{\scriptsize{out}}}\rangle\propto \sum_{u\in {\cal V}} |u\rangle |\psi_u\rangle.$ Here the sum is over all $n$-bit strings $u$ that are orthogonal to $a$ (w.r.t. modulo-2 arithmetic). We denote by ${\cal V}$ the subspace over $\mathbb{Z}_2$ of all such $u$. The $|\psi_u\rangle$ are (irrelevant) normalized states.
Next, all qubits in the first register are measured in the computational basis, yielding a bit string $u$ which is drawn uniformly at random from the subspace ${\cal V}$. Running this procedure $N$ times, one generates the $(Nn)$-qubit state $|\psi_{\mbox{\scriptsize{out}}}\rangle^N$ and one subsequently obtains $N$ bit strings $u^1, \dots, u^N$, each drawn randomly from ${\cal V}$. We assemble these vectors as the rows as an $N\times n$ matrix, denoted by $\mathbf{u}$. If $N=O(n)$ then the probability that $u^1, \dots, u^N$ do \emph{not} span the entire space ${\cal V}$ is exponentially small in $n$. In the final step in the algorithm, one uses a classical computer to compute a solution $x$ to the linear system of equations $\mathbf{u}x=0$. More precisely, in the decision problem version of Simon's algorithm, a function $g:\{0, 1\}^{nN}\to\{0, 1\}$ is computed which takes the entries of the matrix $\mathbf{u}$ as input and which outputs 1 if there exists a solution $x$ where the first bit of $x$ is equal to 1; the output is zero otherwise. Note that $g$ is efficiently computable classically. If the matrix $\mathbf{u}$ has rank $n-1$---which happens in all cases except for an exponentially small fraction---then there is a unique nontrivial solution i.e. $x=a$, in which case the function $g(\mathbf{u})$ correctly outputs the first bit of $a$.

In summary, Simon's algorithm can be implemented with an $(Nn)$-qubit circuit (where $Nn=$ poly$(n)$) displaying the following structure; the circuit acts on the all-zeroes computational basis state.

\begin{itemize}
\item[] {\bf Round  1}: apply a Hadamard gate to some subset of qubits;

{\bf Round  2}: apply an efficiently computable basis-preserving unitary operation;

{\bf Round  3}: apply another round of Hadamard gates to some subset of the qubits; the latter subset is denoted by $S$;

{\bf Round  4}: perform a computational basis measurement on all qubits in $S$. Denote by  $\mathbf{u}$ the bit string containing all measurement outcomes.

{\bf Round  5}: classically compute the value $g(\mathbf{u})$---which represents the output of the algorithm---where $g$ is some efficiently computable Boolean function.
\end{itemize}
For the time being,  we will consider  the above class of 5-round circuits in full generality, and ignore the specific forms of e.g. the functions $f$ and $g$ needed in Simon's algorithm.

\subsubsection{Intermezzo: learning theory}\label{sect_fourier}

In order to formally state and prove theorem \ref{thm_postproc}, beforehand we briefly need to discuss some elementary concepts related to learning theory of Boolean functions (see e.g. \cite{Ma94}). Readers familiar with these concepts may immediately skip to section \ref{sect_proof_thm2}.

\begin{itemize}
\item[1.] A Boolean function is any function $g:\{0, 1\}^m\to \{0, 1\}$. Every Boolean function can be written in a unique way as a multivariate polynomial $g(x) = \sum_S a_S x^S$ over $\mathbb{Z}_2$. In this expression, the sum ranges over all subsets $S\subseteq\{1, \dots, m\}$. Moreover one has $a_S\in\mathbb{Z}_2$ and $x^S:=\prod_{i\in S} x_i$ for every $S$, and arithmetic is performed over $\mathbb{Z}_2$. The ($\mathbb{Z}_2$-)degree of $g$ is the size of the largest set $S$ such that $a_S=1$.

\item[2.] The Fourier  transform $\hat g:\{0, 1\}^m\to\mathbb{R}$ of $g$ is defined as follows: \be \hat g(u) = \sum_{x} (-1)^{u^Tx + g(x)},\ee for every $m$-bit string $u$.  The quantities $\hat g(u)$ are called the Fourier coefficients of $g$. If the function $g$ is computable in poly-time (or provided as an oracle), and if a bit string $u$ is provided as an input, then there exists an elementary poly-time classical algorithm to estimate the quantity $2^{-m}\hat g(u)$ with polynomial accuracy. To see this, simply note that $2^{-m}\hat g(u)$ coincides with the expectation value of the (efficiently computable) function $x\to (-1)^{g(x)+ u^Tx}$ w.r.t. the uniform distribution, such that  a polynomial approximation of $2^{-m}\hat g(u)$ can be achieved in poly-time due to the Chernoff-Hoeffding bound.

\item[3.] A Boolean function is said to be $s$-sparse  if it has precisely $s$ nonzero Fourier coefficients. It is easily verified that every linear function is $1$-sparse. Also, it has been shown that every Boolean function corresponding to a polynomial of degree $d$ is at least $2^d$-sparse \cite{Be99}. In this sense the sparseness of a Boolean function is an indication of its nonlinearity, since high-degree polynomials necessarily have many nonzero Fourier coefficients \cite{foot5}. A (family of) function(s) $g$ is simply called `sparse' if its sparseness satisfies $s\leq$ poly$(m)$.

\item[4.] Interestingly, there exists an efficient algorithm to determine all Fourier coefficients of $g$ that are greater than a given threshold value, in the following sense:

\begin{lem}\cite{Ku93}\label{lem_theta} Suppose that one has access to an oracle computing a Boolean function $g$. Let $p(m)$ denote an arbitrary polynomial in $m$. Then there exists a poly-time algorithm that outputs a collection of $m$-bit strings ${\cal T}\subseteq\{0, 1\}^m$ of size poly$(m)$ containing all $u$ such that $2^{-m}|\hat g(u)|\geq (p(m))^{-1}$.
\end{lem}

Together with the remark made in 2, it follows that there exists a poly-time algorithm that outputs the set ${\cal T}$ together with polynomial approximations of all the quantities $2^{-m}\hat g(u)$, for every $u\in{\cal T}$.
Note that lemma \ref{lem_theta} is a nontrivial result: indeed, a priori it is not obvious that the coefficients $\hat g(u)$ that lie above a certain threshold can be determined efficiently,  since in principle there is an exponentially large space of bit strings $u$ to be searched.

\end{itemize}

\subsubsection{Proof of theorem 2}\label{sect_proof_thm2}

We are now in a position to formally state theorem 2:

\

\noindent {\bf Theorem 2} {\it Consider a quantum circuit displaying the 5-round structure as in section \ref{sect_review_simon}. If the function $g$ computed in the round of classical postprocessing is promised to be sparse, then the entire circuit can be simulated efficiently classically, independent of the specific forms of the other rounds.}

\

\noindent An important ingredient in the proof of theorem 2 will be the $m$-qubit operator $W_g$ (where $m$ denotes the number of bits on which $g$ acts) defined by \be\label{W_matrixelements} \langle u|W_g|v\rangle = 2^{-m} \hat g(u+v) \quad\mbox{ for every } u, v\in\{0, 1\}^m.\ee  Note that each row and each column of $W_g$ contains precisely $s$ non-zero entries, where $s$ is the sparseness of $g$; in other words, {\it the Boolean sparseness of $g$ and the sparseness of the operator $W_g$ coincide}. This correspondence prompts the question of when the operator $W_g$ is \emph{efficiently computable} sparse. It can easily be seen that $W_g$ is ECS if and only if (i) $g$ is sparse and (ii)  there exists an efficient algorithm to determine all those strings $u$ such that $\hat g(u)\neq 0$ \emph{and} the values of the corresponding coefficients $\hat g(u)$. Note however, that finding all $u$ such that $\hat g(u)\neq 0$ is highly nontrivial since some of the non-zero Fourier coefficients may be exponentially small, yet nonzero. Moreover, for general (efficiently computable) $g$ the problem of computing $\hat g(u)$ with exponential precision is $\#$P-hard.  Therefore,  requiring $W_g$ to be ECS is highly stringent.

Fortunately, for our purposes the relevant question will be when $W_g$ can be well-approximated by an ECS operation $A$ with polynomial accuracy; moreover, $A$ itself need not be ECS in the exact sense, but poly-ECS as discussed in the remark below theorem \ref{thm_main}---these are much less stringent demands.  The problem of approximating $W_g$ by such an $A$ is actually possible \emph{for every sparse function $g$}. This is shown in the following lemma; the proof relies on lemma  \ref{lem_theta}.
\begin{lem}\label{lem_last}
Let $g$ be a sparse Boolean function acting on $m$ bits that is provided as an oracle, let the operator $W_g$ be defined as in (\ref{W_matrixelements}) and let $p(m)$ be an arbitrary polynomial. Then there exists a poly-time classical algorithm that outputs a poly-ECS $m$-qubit operation $A$ such that $\| W_g - A\|\leq p(m)^{-1}$.
\end{lem}
{\it Proof: } Let $s\leq$ poly$(m)$ denote the sparseness of $g$. Let $\theta>0$ and let $W_g^{\theta}$ denote the matrix obtained by replacing all entries of $W_g$ that are smaller in absolute value than $\theta$, by zero. That is: $\langle u|W_g^{\theta}|v\rangle$ is equal to $2^{-m} \hat g(u+v)$ if  $|2^{-m} \hat g(u+v)|\geq \theta$, and zero otherwise. For now, $\theta$ is arbitrary but below we will choose $\theta$ to be a suitable polynomial in $m$. Since $W_g$ is $s$-sparse, the matrices $W_g^{\theta}$ and $W_g - W_g^{\theta}$ are $s$-sparse as well. Due to lemma \ref{lem_theta} and the remark below it, for every $\theta = 1/$poly$(m)$, the operator $W_g^{\theta}$ is poly-ECS. Next we show that $\theta$ can be tuned appropriately such $\| W_g - W_g^{\theta}\|\leq p(m)^{-1}$ is satisfied. To do so, let $\| \cdot\|_r$ ($\|\cdot \|_{c}$) denote the maximum row (column) sum norm \cite{foot6}; these norms are related to the spectral norm $\| \cdot\|$ via the inequality $\| X\|^2 \leq \| X\|_r \| X\|_{c}$ for every matrix $X$ \cite{Ho90}. As the matrix $W - W_g^{\theta}$ is $s$-sparse and as every entry of this matrix is at most $\theta$ in absolute value, it holds that $\| W - W_g^{\theta}\|_r\leq s\theta $ and $\| W - W_g^{\theta}\|_{c}\leq s\theta $, and hence \be \| W - W_g^{\theta}\|^2 \leq \| W - W_g^{\theta}\|_r\| W - W_g^{\theta}\|_{c} \leq (s\theta)^2.\ee By choosing $\theta:= (sp(m))^{-1}$ and setting $A:= W_g^{\theta}$ with this choice of $\theta$, we have found a matrix $A$ satisfying the desired conditions. This completes the proof. \finpr

Lemma \ref{lem_last} will be the key ingredient in the proof of theorem 2, which is provided next.

\

{\it Proof of theorem 2:}  The analysis  will be simplified by  considering a slightly alternative version of the 5-round circuits in question, where now the entire computation is performed coherently and there is only a single measurement at the end of the computation. To achieve this, first one goes through rounds 1-3 as indicated. Second,  the function $\mathbf{u}\to g(\mathbf{u})$ is  computed  coherently on the relevant registers,  realized by a unitary operation $U_g$ mapping $U_g: \ |\mathbf{u}\rangle\to |g(\mathbf{u})\rangle|\xi_{\mathbf{u}}\rangle$ for some (irrelevant)  states $|\xi_{\mathbf{u}}\rangle$ \cite{foot7} . Finally, the first qubit is measured in the computational basis. The overall circuit is denoted by $U_T$. Letting $g$ be an arbitrary sparse function, we thus have to show that there exists an efficient classical algorithm to approximate $\langle Z_1\rangle = \langle\mathbf{0}|U_T^{\dagger}Z_1U_T |\mathbf{0}\rangle$ (where $|\mathbf{0}\rangle = |00\dots\rangle$) with polynomial accuracy. For further reference, we denote by $|\psi_2\rangle$ the state obtained after round 2; furthermore,  ${\cal H}$ denotes the tensor product of Hadamard gates applied in round 3. Moreover, let $p(n)$ denote an arbitrary polynomial in $n$.

First, remark that the state $|\psi_2\rangle$ is CT. Denoting $O:={\cal H} U_g^{\dagger} Z_1 U_g {\cal H}$,  one has $\langle Z_1\rangle = \langle\psi_2|O|\psi_2\rangle$. \emph{It is now crucial to note that $O=W_g$, where  $W_g$ is defined in Eq. (\ref{W_matrixelements})}; this identity can easily be verified. This allows us to invoke lemma \ref{lem_last}, yielding in poly-time a poly-ECS operation $A$ satisfying $\| W_g - A\|\leq p(n)^{-1}$. Since $A$ is poly-ECS and since $|\psi_2\rangle$ is CT, according to theorem \ref{thm_main} (cf. also the remark below it) it is possible to approximate $\langle\psi_2|A|\psi_2\rangle$ with polynomial accuracy in poly-time with classical means. In particular, it is possible to efficiently generate a number $c$ such that $|c - \langle\psi_2|A|\psi_2\rangle |\leq p(n)^{-1}$. Since $\langle Z_1\rangle = \langle\psi_2| W_g|\psi_2\rangle$, we then have \be |c - \langle Z_1\rangle| &\leq& |c - \langle\psi_2|A|\psi_2\rangle | + |\langle\psi_2|(W_g -A)|\psi_2\rangle |\nonumber\\&\leq& p(n)^{-1} + \| W_g - A\|\leq 2p(n)^{-1}.\ee In the first inequality we have used the triangle inequality; in the second inequality we have used that $|\langle\psi_2|(W_g -A)|\psi_2\rangle |$ is not greater $\| W_g - A\|$; in the third inequality, we have used that $\| W_g - A\|\leq p(n)^{-1}$. This hence shows that a polynomial approximation of $\langle Z_1\rangle$ can be achieved in poly-time, thus proving the claim. \finpr

We now specialize the discussion to Simon's algorithm. Note that the classical postprocessing in this algorithm is particularly simple, as it merely involves solving a system of linear equations over $\mathbb{Z}_2$. Nevertheless, the function $g$ needed in Simon's algorithm is highly non-sparse. The intuition of the argument is that the function $g(\mathbf{u})$ is related to the computation of the determinant of a suitable matrix (or an analogous function in the case of non-square matrices), since the function $g$ decides whether there exists a nontrivial solution to a certain system of linear equations. It is known that the determinant function $X\to $ det$(X) $ corresponds to a polynomial of degree $k$ in the case of $k\times k$ matrices $X$, i.e. the degree of the polynomial is the square root of the input size $k^2$ of the determinant function. As the degree of a polynomial provides a lower bound to the logarithm of the sparseness (see point 3 in section \ref{sect_fourier}), it follows that the determinant function has exponentially high sparseness $s\geq 2^k$. An analogous argument can be used to show that the function $g$ considered in Simon's algorithm has high sparseness parameter $s$.

Looking at the problem differently, one can in fact use the proved $O(2^{\frac{n}{2}})$ classical oracle lower bound for Simon's problem to immediately \emph{infer} that the function $g$ \emph{cannot be sparse}. Indeed, if $g$ were sparse then our classical simulation results would imply the existence of a classical algorithm to solve Simon's problem using poly$(n)$ classical oracle queries, which is provably not possible. Note that it is remarkable that the classical query lower bound for the oracle $f$ can hence be used to infer properties of another function $g$!

\section{Matchgates and poly-time classical computation}\label{sect_final_byproduct}

We conclude this paper with a result regarding the computational power of matchgate circuits. While seemingly disconnected from the rest of the paper, this result will actually follow from our discussion of Simon's algorithm.

Call a family of functions $f_n:\{0, 1\}^n\to\{0, 1\}$ \emph{efficiently matchgate-computable} if there exists a family of nearest-neighbor matchgate-circuits $U_n$  acting on $M_n=$ poly$(n)$ qubits ($n=1, 2, \dots)$, such that $U_n$,  acting on $|x\rangle|0\rangle^{M-n}$ and followed by a $\{|0\rangle, |1\rangle\}$ measurement on the first qubit, yields the output $f(x)$ with probability $p\geq 2/3$, for all $n$-bit strings $x$. Here $U_n$ is to depend only on the input size $n$ and not on the entire input $x$ (this aspect is important, as will be highlighted in the proof of theorem 2). Moreover,  the circuit family is to be poly-time uniformly generated in the sense that the description of $U_n$ is to be poly-time computable from the number $n$. Our result is the following.

\begin{thm}\label{thm_mg}
There exist functions that are efficiently computable classically (i.e. functions in P) that are not efficiently matchgate-computable.
\end{thm}
An interesting feature of this  result is its proof method. Surprisingly, the proof  will follow from our analysis of Simon's algorithm---even though the latter seems to have nothing to do with matchgates! Roughly speaking, we will show that if theorem \ref{thm_mg} were \emph{false}, then there would exist a quantum circuit to solve Simon's problem that can be simulated classically with our methods---hence resulting in a \emph{classical} algorithm for Simon's problem that requires only poly$(n)$ queries to the oracle. As the latter has been proved to be an impossibility, this will show that theorem \ref{thm_mg} has to be true.

In the proof of theorem \ref{thm_mg} we will need the following simple application of corollary \ref{thm_comp1}.

\begin{itemize}
\item[]{\bf Fact 1:} Consider an $n$-qubit quantum circuit $V=V_4V_3V_2V_1$ where both $V_1$ and $V_3$ represent  collections of Hadamards applied to subsets of the qubits, $V_2$ is  efficiently computable basis-preserving, and $V_4$ is a poly-size (nearest-neighbor) matchgate circuit. Then any such circuit (acting on $|0\rangle^n$ and followed by measurement of $Z_1$) can be simulated
efficiently classically due to corollary \ref{thm_comp1}, taking $V_2V_1\equiv U_1$ and $V_4V_3\equiv U_2$. Indeed, $V_2V_1|0\rangle^n$ is CT and $(V_4V_3)^{\dagger} Z_1(V_4V_3)$ is a linear combination of poly$(n)$ Pauli products and hence ECS.
\end{itemize}

{\it Proof of theorem \ref{thm_mg}:}  Consider the following variant $\tilde g$ of the function $g$  computed in the classical postprocessing in Simon's algorithm: $\tilde g$ takes an $N\times n$ matrix $\mathbf{u}$ together with an integer $i$ between 0 and $n$ (specified in terms of $\log n$ bits) as its inputs, and outputs 1 if and only if there exists a bit string $x=(x_1, \dots, x_n)$ satisfying $\mathbf{u}x=0$ and $x_i=1$. Note that $\tilde g$ is efficiently computable classically. We claim that $\tilde g$ is \emph{not} efficiently matchgate-computable. To prove this, we show that the converse leads to a contradiction. Suppose that $\tilde g$ is matchgate-computable and let $U$ denote the (family of) matchgate circuit(s) that computes $\tilde g$.  Now consider the following quantum algorithm ${\cal A}$:  first prepare the state $|i\rangle\otimes|\psi_{\mbox{\scriptsize{out}}}\rangle^{\otimes N}$, where $N=O(n)$ and where $|\psi_{\mbox{\scriptsize{out}}}\rangle\propto \sum_{u\in {\cal V}} |u\rangle |\psi_u\rangle$  as in section \ref{sect_review_simon}; up to a permutation of the qubits, at this point the state of the quantum register has the form $\sum_{\mathbf u} |i\rangle |\mathbf{u}\rangle|\chi_{\mathbf{u}}\rangle$ for some (irrelevant) normalized $|\chi_{\mathbf{u}}\rangle$, and where the sum is over all $N\times n$ matrices $\mathbf{u}$ for which each row belongs to ${\cal V}$. Second, apply the matchgate circuit $U$ on the relevant registers in order to compute $|i, \mathbf{u}\rangle\to |\tilde g(i, \mathbf{u})\rangle$ in superposition---note that at this point it is crucial that $U$ only depends on the input size but not on the entire input. Finally, measure $Z_1$ and let $\langle Z_1\rangle$ denote the expectation value of $Z_1$. If the $i$-th bit of the unknown string $a$ in Simon's problem is equal to 0, then $\langle Z_1\rangle$ lies exponentially close to 1/3;  if this bit is 1 then  $\langle Z_1\rangle$ lies exponentially close to $-1/3$. It is now easily verified that the  algorithm ${\cal A}$ is implemented with a circuit displaying the structure considered in Fact 1. Hence a polynomial approximation of $\langle Z_1\rangle$ can be classically achieved in poly-time with exponentially small probability of failure, for every $i$.  Note that such an approximation allows to decide whether $\langle Z_1\rangle$ lies exponentially close to 1/3 or $-1/3$. This hence leads to a poly-time \emph{classical} algorithm to determine $a$. This comprises a contradiction, given the $O(2^{\frac{n}{2}})$ classical query lower bound for Simon's problem. Hence, $g$ cannot be efficiently matchgate-computable. \finpr

\subsection*{Acknowledgements}
I am very grateful to R. Jozsa for discussions and suggestions on the manuscript, and to H. Briegel, I. Cirac, W. D\"ur, G. Giedke, B. Kraus, R. Renner, N. Schuch and K. Vollbrecht for discussions. Work supported by the excellence cluster MAP.

\appendix

\section{Appendix: Sampling and the Chernoff-Hoeffding bound}\label{sect_chernoff}
The Chernoff-Hoeffding bound is a tool to assess with which precision the expectation value of a random variable may be approximated in terms of `sample averages'. This bound asserts the following. Let $X_1, \dots X_K$ be i.i.d. real-valued random variables with $E := \mathbb{E}X_i$ and $X_i\in [-1, 1]$ for every $i=1, \dots, K$. Then \be\label{hoeff} \mbox{Prob} \left\{ \left|\frac{1}{K}\sum_{i=1}^K X_i - E\right| \leq \epsilon\right\} \geq 1-2 e^{- \frac{K\epsilon^2}{4}}.\ee In the case of complex-valued random variables $X_i$, a similar bound can be
obtained for $|X_i|\leq 1$ by splitting $X_i$ in its real and imaginary part and using (\ref{hoeff}) on both of these parts.
In this work we will consider the Chernoff-Hoeffding bound in the following context. Let ${\cal P}:=\{p_x\}$ be a probability distribution on the set of $n$-bit strings $x\in\{0, 1\}^n$ and let $x\to F(x)\in\mathbb{C}$ be a complex function such that $|F(x)|\leq 1$ for every $x$.  Let $\langle F\rangle= \sum_x p_x F(x)$ denote the expectation value of $F$. The goal is to approximate $\langle F\rangle$ by sampling from the
distribution ${\cal P}$. To do so, consider $K$ $n$-bit strings $x^1, \dots, x^K$ drawn (independently) from the distribution ${\cal P}$, and denote the average $\sigma:= K^{-1} \sum_{i=1}^K F(x^i).$ The Chernoff-Hoeffding bound then implies the following. For every $\epsilon = p(n)^{-1}$, where $p(n)$ represents an arbitrary polynomial in $n$, there exists a $K$ that scales at most polynomially with $n$, such that the inequality $|\sigma- \langle F\rangle|\leq \epsilon$ holds with a probability that is exponentially (in $n$) close to 1. 
In other words, by taking poly$(n)$ samples $x^i$ it is possible to estimate $\langle F\rangle$ with an error that scales as $p(n)^{-1}$ for every choice of  $p(n)$. We will henceforth denote this type of estimate as an approximation with `polynomial accuracy' or a `polynomial approximation'. Note that a polynomial approximation achieves an estimate of $\langle F\rangle$ up to $O(\log n)$ significant bits.

Moreover, if the function $F$ can be evaluated in poly-time \emph{and} if it is possible to   sample in poly-time from ${\cal P}$, then the quantity $ \sigma$ can be computed in poly-time. Hence,  an overall efficient method is achieved to compute a polynomial approximation of $\langle F\rangle$  with exponentially small probability of failure.  In this paper we will mostly ignore the fact that the Chernoff-Hoeffding bound yields polynomial approximations that do not succeed with unit probability but rather with a probability that is exponentially close to one. When the notion of a polynomial approximation is considered in the text, we will mean a polynomial approximation that is achieved with a probability that is exponentially close to one.

We discuss two immediate generalizations of the above arguments. First, above we have required that the function  $F$ can be evaluated with perfect precision in poly-time. Such perfect accuracy is in this context not necessary. In particular, with similar methods as above, a polynomial approximation of $\langle F\rangle$ can be achieved in poly-time if \emph{$F(x)$ itself can be approximated with polynomial accuracy  in poly-time}. This can be seen as follows. Suppose that, on input of an arbitrary $x$, a polynomial approximation of $F(x)$ can be achieved in poly-time. Let $p(n)$ be an arbitrary polynomial and consider $K$ $n$-bit strings $x^1, \dots, x^K$ drawn from the distribution ${\cal P}$ as before. Then for large enough $K$ (where $K$ scales as a polynomial in $n$ with suitably high degree), $K^{-1} \sum_{i=1}^K F(x^i)$ lies $\epsilon$-close to $\langle F\rangle $, where $\epsilon=(2p(n))^{-1}$. As each of the $K$ quantities $F(x^i)$ can be approximated with polynomial accuracy in poly-time by assumption, it is possible to efficiently generate $K$ complex numbers $c^i$ $(i=1, \dots, K)$ such that $|c^i - F(x^i)|\leq (2p(n))^{-1}$. Using the triangle inequality and denoting $c:= K^{-1} \sum_{i=1}^K c^i$, it then easily follows that  $|\langle F\rangle - c|\leq p(n)^{-1}.$

Second, so far we have considered functions $F$ satisfying $\|F\|:=\max_x |F(x)|\leq 1$. Note that similar conclusions can be reached for functions satisfying $\| F\|\leq $ poly$(n)$.

The discussion in the present section can be summarized as follows.
\begin{thm}[Chernoff-Hoeffding bound]\label{thm_chernoff}
Suppose that it is possible to sample in poly-time with classical means from a probability distribution $\{p_x\}$ on the set of $n$-bit strings. Let $F:\{0, 1\}^n\to \mathbb{C}$ denote a function satisfying $\|F\|\leq $ poly$(n)$. Moreover, suppose that it is possible to efficiently estimate $x\to F(x)$ with polynomial accuracy on a classical computer. Then there exists an efficient classical algorithm to estimate $\langle F\rangle$ with polynomial accuracy.
\end{thm}


\begin{thebibliography}{99}
\bibitem{foot3}
Recall that an $n$-qubit Pauli operator (or Pauli product) has the form $P=P_1\otimes\dots\otimes P_n$, where each $P_i$ is either the $2\times 2$ identity or one of the Pauli matrices $X$, $Y$ or $Z$.
\bibitem{foot4''}
A function $f$ is constant if $f(x)=f(0)$ for every $n$-bit string $x$; $f$ is balanced if exactly $2^{n-1}$ inputs $x$ satisfy $f(x)=0$.
\bibitem{foot5}
The converse, however, is not true. For example, there exist degree-2 polynomials with sparseness $s=2^n$ (the inner product function is an example of this).
\bibitem{foot6}
That is, $\| X\|_r: = \max_i \sum_{j=1}^N |X_{ij}|$ and $\| X\|_{c}: = \max_j \sum_{i=1}^N |X_{ij}|$, for every $N\times N$ matrix $X$.
\bibitem{foot7}
We remark that in  the definition of $U_g$, in the most general case one must allow $U_g$ to use $m=$ poly$(n)$ ancillary qubits prepared in, say, the state $|0\rangle$, i.e. $U_g: \ |\mathbf{u}\rangle|0\rangle^m\to |g(\mathbf{u})\rangle|\xi_{\mathbf{u}}\rangle$. For clarity, we have not incorporated this in the proof; the argument can be generalized appropriately without significant complications.
\end{thebibliography}
\end{document}